\documentclass[aps,prd,twocolumn,nofootinbib]{revtex4-1}
\usepackage{epsfig}
\usepackage[colorlinks,linkcolor=blue,anchorcolor=blue,citecolor=blue,urlcolor=blue,breaklinks=true]{hyperref}
\usepackage{amsmath}
\usepackage{booktabs} 
\usepackage{multirow}
\usepackage{array}
\usepackage{graphics}
\usepackage{slashed}
\usepackage{color}
\usepackage{amsmath}
\usepackage{bm}
\usepackage{setspace}
\usepackage{comment}
\emergencystretch=\maxdimen
\begin{document}
\author{He-Xia Zhang}\email{hexiazhang@mails.ccnu.edu.cn}
\address{ Key Laboratory of Quark \& Lepton Physics (MOE) and Institute of 
	Particle Physics, Central China Normal University, Wuhan 430079, China}
 \author{Ben-Wei Zhang}
\email{bwzhang@mail.ccnu.edu.cn}
\affiliation{Key Laboratory of Quark \& Lepton Physics (MOE) and Institute of  Particle Physics, Central China Normal University, Wuhan 430079, China}
\affiliation{Institute of Quantum Matter, South China Normal University, Guangzhou 510006, China}

\title{The effect of momentum anisotropy on  quark matter in the quark-meson model} 

\begin{abstract}

	We investigate the chiral phase structure of  quark matter with  spheroidal momentum-space anisotropy specified by one anisotropy
	parameter $\xi$ in the 2+1 flavor quark-meson  model. 
	We find  that the chiral phase diagram and  the location of the  critical endpoint (CEP)  are affected significantly by  the value of $\xi$.  With the increase of $\xi$,  the CEP is shifted to smaller temperatures and larger quark chemical potentials. And the  temperature of the CEP is more sensitive to the anisotropy parameter than the corresponding quark chemical potential, which is opposite to the study for finite system volume effect.
   Furthermore,  the effects of momentum anisotropy  on the thermodynamic properties and  scalar (pseudoscalar) meson masses  are  also studied at vanishing quark chemical potential.  The numerical results  show that an increase of  $\xi$ can hinder the restoration of chiral symmetry. We also find  that  shear viscosity and electrical conductivity decrease as $\xi$ grows. However,  bulk viscosity exhibits a significant  non-trivial behavior with $\xi$  in the entire temperature domain of interest.

\bigskip

\end{abstract}

\maketitle

\section{INTRODUCTION}

Quantum chromodynamics (QCD) is the fundamental theory for describing  the strong interaction, and its phase structure is an important  subject of great interest in recent decades.
The  first-principle results from lattice QCD simulation~\cite{lattice1,lattice2} have indicated that  with increasing temperature $T$, the transition  from the ordinary nuclear matter to the chiral symmetric quark-gluon plasma (QGP) is  a smooth crossover  at  small or zero chemical potential $\mu$.
  At large chemical potential, lattice QCD simulation as a reliable tool to obtain the chiral properties of QCD matter, confronts a great challenge  due to the fermion sign problem~\cite{Splittorff:2007ck}, although different strategies (for reviews see, e.g., Refs.~\cite{Braun-Munzinger:2015hba,Fukushima:2013rx,Fukushima:2010bq}), such as Taylor series expansions~\cite{Allton:2005gk,Gavai:2003mf,Gavai:2008zr}, imaginary chemical potential, reweighting techniques~\cite{Fodor:2001pe,Fodor:2002km},~complex Langevin method~\cite{Aarts:2009uq,Klauder:1983sp}, have been developed to try to  tackle this problem. 
 In this context, some alternative theoretical tools, such as  QCD low-energy effective  models (e.g. the Nambu-Jona-Lasinio  model~\cite{Nambu:1961fr,Klevansky:1992qe,Hatsuda:1994pi}, Polyakov-loop extended NJL (PNJL) model~\cite{Meisinger:1995ih,Roessner:2006xn,Ratti:2007jf}, quark-meson  model or linear sigma model~\cite{Schaefer:2006ds,Schaefer:2007pw,Schaefer:2004en,Lenaghan:2000ey,SchaffnerBielich:1999uj}, Polyakov quark-meson (PQM) model~\cite{Schaefer:2011ex,Gupta:2009fg,Schaefer:2009ui,Stiele:2016cfs}),  Dyson-Schwinger equation approach~\cite{Bashir:2012fs,Fischer:2018sdj}, the functional renormalization group approach~\cite{Gies:2006wv,Pawlowski:2005xe,Schaefer:2006sr,Bagnuls:2000ae}, which are not restricted by chemical potential,  have been proposed to better explore the QCD phase structure at high chemical potential.
And the  results from the effective model calculations~\cite{Gupta:2011ez,Schaefer:2008hk}
 show that the chiral phase transition of the strongly interacting matter is a first-order transition  at high density, and  a second-order critical endpoint (CEP) can exist between the crossover line and the first-order phase transtion line in the ($\mu$, $T$)-plane. 
Apart from the phase transition,  other important informations, such as thermodynamic properties,   in-medium properties of mesons\cite{Tawfik:2014gga,Schaefer:2008hk} and  transport properties~\cite{Abhishek:2017pkp,Ghosh:2018xll,Singha:2017jmq} for the strongly interacting matter  are also  extensively studied in  these QCD  effective models.

To take into account the intricacy of the realistic quark matter produced in relativistic heavy-ion collisions (HICs) at the RHIC and the LHC, different improved versions of  the QCD effective models have been proposed by including the effects of  the finite volume of the system~\cite{Saha:2017xjq,Bhattacharyya:2012rp,Zhang:2019gva,Magdy:2019frj,Ya-Peng:2018gkz,Deb:2020qmx,Abreu:2019czp,Shi:2018tsq,Tripolt:2013zfa,Braun:2011iz,Tripolt:2013zfa,Li:2017zny,Palhares:2009tf,Magdy:2015eda,Braun:2005fj,Braun:2010vd,Zhao:2019ruc}, the non-extensive effects in term of  long-distance correlation~\cite{Zhao:2020xob,Shen:2017etj}, the presence of magnetic fields~\cite{Fukushima:2010fe,Andersen:2014xxa,Ruggieri:2013cya,Mao:2016fha,Gatto:2010pt,Fukushima:2010fe,Ghosh:2019lmx,Kashiwa:2011js,Andersen:2014oaa,Yu:2014xoa,Andersen:2013swa},  and the effects of  electric field~\cite{Tavares:2019mvq,Tavares:2018poq,Cao:2015dya,Ruggieri:2016lrn,Ruggieri:2016xww}, to better  explore the chiral/confinement  properties  of the strongly interacting matter at finite temperature or quark chemical potential.
Conventionally, in the literature, all the effective  models or improved effective models are    based on an ideal assumption that the constituents of  quark matter are completely  isotropic in momentum-space  for the absence of    magnetic fields. 
However, due to the  geometry  of fireball created in HICs  is asymmetric, the system evolves with  different pressure gradients along  different directions. As a result, the  expanding and cooling rate along the beam direction (denotes as longitudinal direction) is larger than  radial direction~\cite{Romatschke:2003ms} and this momentum anisotropy  can survive  in all the stages of the HICs,  consequently, the  parton-level momentum distribution functions  may  become anisotropic. Thus, it's essential to  consider  the momentum-space anisotropy induced by the  rapid longitudinal asymptotic  expansion into the phenomenological investigation of  different observables.
Up to present, extensive works have been made to  explore the effects of momentum anisotropy  on  the parton self-energy~\cite{Romatschke:2003ms,Schenke:2006fz,Kasmaei:2018yrr,Kasmaei:2016apv}, photon and dilepton production~\cite{Bhattacharya:2015ada,Kasmaei:2019ofu,Schenke:2006yp,Kasmaei:2018oag}, the dissociation of quarkonium~\cite{Jamal:2018mog,Burnier:2009yu,Thakur:2012eb}, heavy-quark potential~\cite{Dumitru:2007hy,Nopoush:2017zbu}, various  transport coefficients~\cite{Rath:2019vvi,Zhang:2020efz,Thakur:2017hfc,Srivastava:2015via}, jet quenching parameter\cite{Giataganas:2012zy} which,  are sensitive to the evolution of the QGP.
And  associated results  have indicated that  the momentum-space anisotropy has a  significant effect on the observables of the QGP.
However, with the best of our knowledge, so far there is no study of momentum anisotropy in the framework of effective QCD models and no research regarding  the effect of  momentum-space anisotropy on chiral phase transition. Inspired by this fact, one major goal of present work is to reveal  how the  momentum anisotropy  qualitatively affects the chiral phase structure as well as  transport properties  in the strongly interacting matter.

The present paper is a first attempt to study the effect of the momentum-space anisotropy induced by  the rapid longitudinal expansion of fireball created in HICs on  the QCD  chiral phase transition. 
 We adopt the  2+1 flavor  quark-meson model, which is successful in describing the mechanism of spontaneous chiral symmetry breaking, to approximate  quark matter. The effect of  momentum anisotropy  enters in  the  quark-meson model by  substituting  the isotropic  (local equilibrium)   distribution function   in the total thermodynamical potential  with  the anisotropic one.
 This introduces one more degree of freedom, {\it  viz},  the direction of anisotropy. The anisotropic parameter $\xi$,  representing the degree of momentum anisotropy or the tendency of the system to stay away from the isotropic state, is  also considered as argument into the isotropic distribution function.  
 Based on this momentum anisotropy-dependent quark-meson model, we first explore how the momentum anisotropy affects the chiral phase diagram and the location of CEP.  Next, we investigate the  thermodynamic  properties and  the thermal properties of  various  scalar (pseudoscalar) meson masses for vanishing chemical potential in both isotropic and  anisotropic quark matter.
  Finally,  transport coefficients, such as  shear viscosity, electrical conductivity,  and bulk viscosity,  which are crucial to understand the dynamical evolution of QCD matter, also are estimated  in an (an-)isotropic quark matter.
 Note that  we restrict ourselves here to the anisotropic system  close to  isotropic local equilibrium state, consequently, 
 the calculations of thermodynamic quantities, meson masses  and transport coefficients in the  anisotropic system  are   methodologically  similar to those in the isotropic system. 
 Especially, in the small $\xi$ limit, the anisotropic distribution can just  linearly  expand to the linear  order of $\xi$.  Using  this  linear approximation of the anisotropic distribution, 
 the mathematical expression of transport coefficients,  which are obtained by solving the relativistic Boltzmann equation under the relaxation time approximation,  can be explicitly separated into   an equilibrium part and  an anisotropic  correction part~\cite{Rath:2019vvi,Zhang:2020efz,Thakur:2017hfc,Srivastava:2015via}. For $\xi\rightarrow 0$, the analytic expressions can  reduce to   the standard expressions in  the local equilibrium medium, which can be seen in  Section.~\ref{IV}.
 
 This paper is organized as follows. In section.~\ref{QM-Model}, we give a brief overview of the three-flavor quark-meson model.  In section.~\ref{III}, the modification of the thermodynamical potential within momentum-space anisotropy is presented.
 In section.~\ref{IV},  we discuss the chiral phase transition, thermodynamics properties, meson masses, and transport coefficients in both isotropic and anisotropic quark matter.
In section.\ref{summary}, we summarize the main results and give an outlook.

\section{the  quark-meson model}\label{QM-Model}
The quark-meson model as a successful  QCD-like effective model can  capture an important feature of QCD, namely, chiral symmetry breaking and  restoration  at high temperature/density. The Lagrangian of  the  three-flavor quark-meson model presently used for our purpose  is taken from Ref.~\cite{Lenaghan:2000ey}:
\begin{eqnarray}\label{Lagrangian}
\mathcal{L}_{\textrm{QM}} =\bar{\Psi}(i\gamma_{\mu}D^{\mu}-g\phi_5)\Psi+\mathcal{L}_{\mathrm{M}},
\end{eqnarray}
where $\Psi=u,d,s$ stands for the  quark field with three flavors ($N_{f}=3$)  and three color degrees of freedom ($N_{c}=3$). The first term in the right hand  side of  Eq.~(\ref{Lagrangian}) represents  the interaction between the  quark field  and  the scalar ($\sigma$) and pseudoscalar ($\pi$) fields with a flavor-blind Yukawa coupling $g$ of the quarks to the mesons. The  meson matrix is given as 
\begin{eqnarray}
 \phi_5=T_a(\sigma_{a}+i\gamma_5\pi_a),
\end{eqnarray}
where $T_{a}=\lambda_a/2$ with $a=0,\dots,8$ are the nine generators of the U(3) symmetry. $\lambda_{a}$ is Gell-Mann matrix with $\lambda_{0}=\sqrt{\frac{2}{3}}1$. $\sigma_{a}$ and $\pi_a$ denote the scalar meson  nonet and the pseudoscalar meson nonet, respectively.

The second term in Eq.~(\ref{Lagrangian})  is  the purely mesonic contriburion, $\mathcal{L}_{\mathrm{M}}$, which   describes the chiral symmetry breaking parttern in strong interaction.  It is given by~\cite{Lenaghan:2000ey}
\begin{eqnarray}\label{eq:meson}
\mathcal{L}_{\mathrm{M}}&=&\mathrm{Tr}(\partial_\mu\phi^{\dagger}\partial^{\mu}\phi-m^2\phi^{\dagger}\phi)-\lambda_1[\mathrm{Tr}(\phi^{\dagger}\phi)]^2\nonumber\\
&&-\lambda_2\mathrm{Tr}(\phi^{\dagger}\phi)^2+c[\mathrm{Det}(\phi)+\mathrm{Det}(\phi^{\dagger})]\nonumber\\
&&+\mathrm{Tr}[H(\phi+\phi^{\dagger})],
\end{eqnarray}
with $\phi=T_{a}\phi_{a}=T_{a}(\sigma_{a}+i\pi_a)$ representing a complex ($3\times3$)-matrix. Explict chiral symmetry breaking is shown in the last term of Eq.~(\ref{eq:meson}), where $H=T_{a}h_{a}$  is a ($3\times3$)-matrix with nine external fields $h_{a}$. Explict $U(1)_{A}$ symmetry is given by  't Hooft determinant term with the anomaly term $c$.  $m^2$	is the tree-level mass of the fields in the absence of symmetry breaking, $\lambda_1$ and $\lambda_2$ are the two possible quartic coupling constants.

\begin{center}
	\begin{table}
		\caption{The parameters  used in our work from Ref.~\cite{Schaefer:2008hk}.}\label{tb1}
			\begin{tabular}{cccccc}
			\hline\hline
			$m^2[\mathrm{MeV^2}]$&$h_x[\mathrm{MeV^3}]$&$h_y[\mathrm{MeV^3}]$&$\lambda_1$&$\lambda_2$&$c[\mathrm{MeV}]$\\
			\hline
			$(342.252)^2$&$(120.73)^3$&$(336.41)^3$&1.4&46.68&4807.84\\
			\hline\hline
		\end{tabular}
	\end{table}
\end{center}

Under the mean-field approximation~\cite{Schaefer:2008hk}, the total thermodynamic potential density of  the 
quark-meson model at finite temperature $T$ and quark chemical potential $\mu_f$ is given by 
\begin{eqnarray}\label{eq:4}
\Omega(T,\mu_f)=
\Omega_{q\bar{q}}(T,\mu_f)+U(\sigma_{x},\sigma_{y})_{}.
\end{eqnarray}
The first term  $\Omega_{q\bar{q}}$ in the right hand of Eq.~(\ref{eq:4}) denotes  the fermionic part of the thermodynamic potential~\cite{Schaefer:2008hk}:
\begin{eqnarray}\label{potential}
\Omega_{q\bar{q}}(T,\mu_f)&=&2N_{c}\sum_{f=u,d,s}^{}T\int_{}^{}\frac{{\rm d}^3\mathbf{p}}{(2\pi)^3}[\ln(1-f_{q,f}^0(T,\mu_{f},\mathbf{p}))\nonumber\\
&&+\ln(1-f_{\bar{q},f}^0(T,\mu_{f},\mathbf{p}))],
\end{eqnarray}
with  the isotropic  equilibrium distribution function of (antiquark) quark for $f$-th flavor
\begin{equation}
f^0_{q(\bar{q}),f}(T,\mu_{f},\mathbf{p})=\frac{1}{\exp[E_{f}\mp\mu_{f}/T]+1}.
\end{equation}
Here, $E_{f}=\sqrt{p^{2}+m_f^{2}}$ is the single-particle  energy with flavor-dependent  constituent quark mass $m_{f}$. The sign $\mp$ corresponds to quarks and antiquarks, respectively.
In present work, an uniform quark  chemical potential $\mu\equiv\mu_{u}\equiv\mu_{d}\equiv\mu_{s}$ is assumed. 
And the breaking of the $SU(2)$ isospin symmetry is not considered, consequently, the up and down quarks have approximately the same masses, i.e., $m_{u}\approx m_{d}$. In the  quark-meson model, the constituent quark masses are given as
\begin{equation}
m_{l}=g\sigma_x/2\quad \mathrm{and}\quad m_{s}=g\sigma_y/\sqrt{2},
\end{equation}
where $l$ denotes light quarks ($l\equiv u,d$).  $\sigma_x$ and $\sigma_y$ stand for the non-strange and strange chiral condensates, respectively.
The Yukawa coupling $g$ is fixed  to reproduce a light constituent quark mass of $m_l\approx300$~MeV. 
 The second term $U(\sigma_{x},\sigma_{y})_{}$, $viz$, the purely mesonic potential, is given as~\cite{Schaefer:2004en,Lenaghan:2000ey,Schaefer:2011ex}
\begin{eqnarray}
U &=&-h_x
\sigma_x-h_y \sigma_y+ \frac{m^2(\sigma^2_x+\sigma^2_y)}{2} -\frac{c\sigma^2_x \sigma_y }{2\sqrt{2}} \nonumber \\
&&+ \frac{\lambda_1 \sigma^2_x \sigma^2_y}{2} +\frac{(2 \lambda_1
	+\lambda_2)\sigma^4_x}{8}  + \frac{ (\lambda_1+\lambda_2)\sigma^4_y}{4}, \nonumber\\
\end{eqnarray}
where the model parameters: $m^2$, $h_x$, $h_y$, $\lambda_1$, $\lambda_2$ and $c$  as reported in Ref.~\cite{Schaefer:2008hk},  are shown in Table~\ref{tb1}.
Finally, the behaviors  of $\sigma_{x}$ and $\sigma_{y}$ as the  functions of temperature and quark chemical potential can be obtained  by minimizing the total thermodynamic potential density, i.e.,
\begin{eqnarray}\label{eq:gap}
\frac{\partial\Omega}{\partial \sigma_{x}}=\frac{\partial\Omega}{\partial \sigma_{y}}\bigg|_{\sigma_{x}=\bar{\sigma}_x,\sigma_{y}=\bar{\sigma}_y}=0,
\end{eqnarray}
with $\sigma_{x}=\bar{\sigma}_x, \sigma_{y}=\bar{\sigma}_y$ being the global minimum.
\section{Thermodynamic potential with momentum anisotropy}\label{III}

Due to the rapid  longitudinal  expansion of the partonic matter created in the HICs,   an anisotropic deformation of the argument of the isotropic (equilibrium)  parton distribution  functions is generally used to simulate the momentum anisotropy of  QGP
~\cite{Romatschke:2003ms,Schenke:2006fz,Kasmaei:2018yrr,Kasmaei:2016apv,Bhattacharya:2015ada,Kasmaei:2019ofu,Schenke:2006yp,Kasmaei:2018oag,Jamal:2018mog,Burnier:2009yu,Thakur:2012eb,Dumitru:2007hy,Nopoush:2017zbu,Rath:2019vvi,Zhang:2020efz,Thakur:2017hfc,Srivastava:2015via}.
A special and widely used spherical momentum deformation introduced by Romatschke and Strickland ~\cite{Romatschke:2003ms}, which is characterized  by removing and adding particles along a single momentum anisotropy direction, is applied in this paper.
 Accordingly, the local distribution function of $f$-th flavor quarks(antiquarks)  in an anisotropic system can be obtained from the isotropic (local equilibrium) distribution function by the rescaling of one preferred direction in momentum space, which is given as
\begin{eqnarray}\label{eq: fx}
f_{aniso}^{0}(T,\mu_{f},\mathbf{p})=\frac{1}{e^{(\sqrt{\mathbf{p}^2+\xi(\mathbf{p}\cdot\mathbf{n})^2+m_f^2}\mp\mu_{f})/T}+1},
\end{eqnarray}
Here, the anisotropy parameter  $\xi$,  presenting the degree of momentum-space anisotropy,  generally can be defined as
\begin{eqnarray}
\xi=\frac{\left\langle\mathbf{p}_{T}^2\right\rangle}{2\langle p_{L}^2\rangle}-1,
\end{eqnarray}
where $p_{L}$ and $\mathbf{p}_{T}$ are the components of  momentum parallel and perpendicular to the direction of anisotropy, $\mathbf{n}$, respectively.
And $\mathbf{p}=(p\sin\theta\cos\phi,p\sin \theta\sin\phi,p\cos\theta)$, where we use a notation $|\mathbf{p}|\equiv p$ for convenience.  $\mathbf{n}=(\sin\alpha,0,\cos\alpha)$, $\alpha$ is the angle between $\mathbf{p}$ and $\mathbf{n}$. Accordingly, $(\mathbf{p}\cdot\mathbf{n})^2=p^2(\sin\theta\cos\phi\sin\alpha+\cos\theta\cos\alpha)^2=p^2c(\theta,\phi,\alpha)$.
 Note that  $\xi>0$  corresponds to a contraction of the particle distribution in the direction of anisotropy whereas $-1<\xi<0$ stands for a stretching of the particle distribution in the direction of anisotropy.

If the system  is close to the ideal massless parton  gas and $\xi$ is small, $\xi$ is also related to the ratio of  shear viscosity to entropy density $\eta/s$  as well as the proper time $\tau$ of the  medium. The relation for one-dimensional  Bjorken expansion  in the Navier-Stokes limit is given as~\cite{Asakawa:2006jn}
\begin{eqnarray}
\xi=\frac{10}{T\tau}\frac{\eta}{s}. 
\end{eqnarray} 
This implies that non-vanishing  shear viscosity combined with  finite momentum relaxation rate in an expanding system   can also  contribute to the momentum-space anisotropy.  At the RHIC energy with the critical temperature $T_{c}\approx160$~MeV, $\tau\approx6 $ fm/c and $\eta/s=1/4\pi$, we can obtain $\xi\approx 0.3$.

In this work, we assume the system has  a small deviation  from the mometum-space isotropy, therefore the value of $\xi$ is small ($|\xi|\ll1$) and
the  Eq.~(\ref{eq: fx}) can be expanded up to linear order in $\xi$,
\begin{eqnarray}\label{eq:f_aniso}
f_{aniso}^0(\mathbf{p})&\approx&f^0_{q,f}-\frac{\xi(\mathbf{p\cdot\mathbf{n}})^2}{2E_{f}T}e^{(E_{f}-\mu_{f})/T}f_{q,f}^{02}\nonumber\\
&=&f_{q,f}^0-\frac{\xi(\mathbf{p\cdot\mathbf{n}})^2}{2E_{f}T}f_{q,f}^{0}(1-f_{q,f}^0).
\end{eqnarray}
By replacing the isotropic distribution functions in Eq.~(\ref{potential})  with the Eq.~(\ref{eq:f_aniso}), we finally obtain  the $\xi$-dependent thermodynamic potential density of fermionic part
\begin{eqnarray}\label{potential-xi}
\begin{aligned}
&\Omega_{q\bar{q}}=
2N_{c}\sum_{f}^{}\int_{}^{}\frac{T{\rm d}^3\mathbf{p}}{(2\pi)^3}\nonumber\\
&\left\{\ln(1-f_{q,f}^0+\frac{\xi p^2c(\theta,\phi,\alpha)}{2E_{f}T}f_{q,f}^{0}(1-f_{q,f}^0))\right.\\
&\phantom{=\;\;}\left.+\ln(1-f_{\bar{q},f}^0+\frac{\xi p^2c(\theta,\phi,\alpha)}{2E_{f}T}f_{\bar{q},f}^{0}(1-f_{\bar{q},f}^0))\right\}.
\end{aligned}\nonumber\\
\end{eqnarray}
 Similar to the studies regarding finite-size effect~\cite{Saha:2017xjq} and non-extensive effect~\cite{Zhao:2020xob}, we also  treat  the anisotropy parameter $\xi$ as a thermodynamic argument in the same footing as $T$ and $\mu$, and   do not have any modifications to the usual quark-meson model parameters due to the presence of  momentum anisotropy. Replacing  the fermionic thermodynamic potential  in Eq.~(\ref{eq:gap}) with Eq.~(\ref{potential-xi}), we can finally   obtain the $\xi$-dependent chiral condensates at finite temperature and quark chemical potential.
\begin{center}
	\begin{table}
		\caption{The chiral critical temperature of the non-strange  condensate $T^{\chi}_c$ and strange  condensate $T^{\chi}_s$  at vanishing quark chemical potential for different anisotropy parameters.}\label{tb2}
		\begin{tabular}{ccccc}
			\hline\hline
			$ \xi $&$-0.4$&$0$&$0.2$&$0.4$\\
			\hline
			$T^{\chi}_c(\mathrm{MeV})$&137&146&152&159\\
			\hline
			$T^{\chi}_s(\mathrm{MeV})$&233&248&258&270\\
			\hline\hline
		\end{tabular}
	\end{table}
\end{center}

\section{results and discussions}\label{IV}
\subsection{phase transition and phase diagram}
\begin{figure}
	\includegraphics[width=0.47\textwidth]{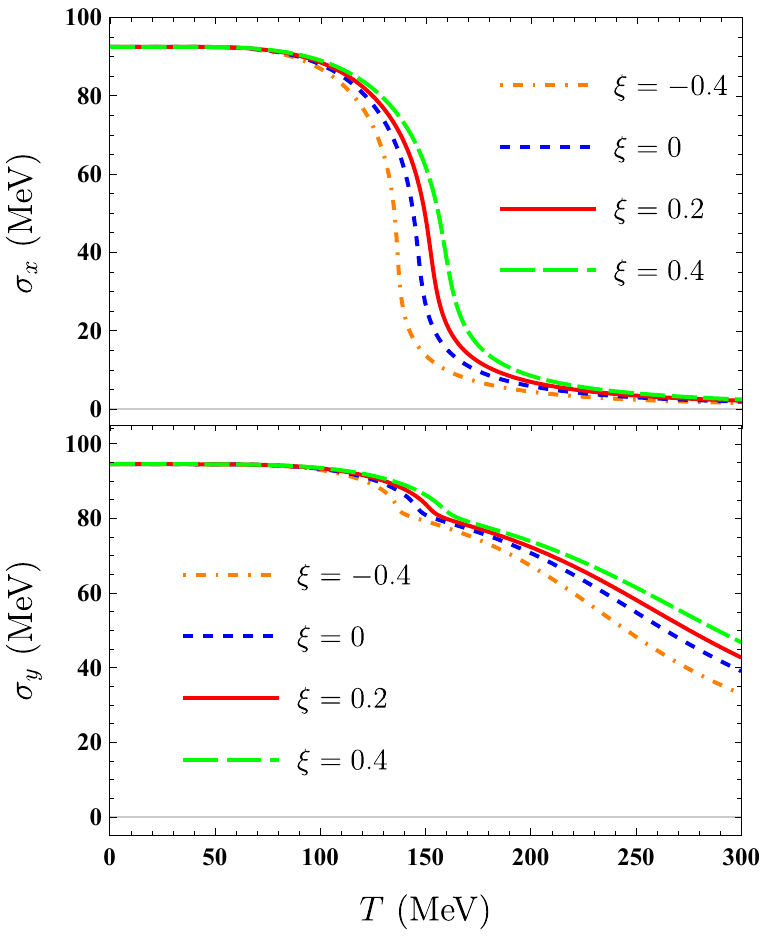}
	\caption{\label{Fig:Condensate}The temperature dependences of non-strange chiral condensate $\sigma_{x}$  (upper panel) and strange  chiral condensate $\sigma_{y}$  (lower panel) at vanishing quark chemical potential for both  isotropic ($\xi=0$ (blue dashed lines))  and anisotropic (i.e., $\xi=$ $-0.4$ (orange dotted-dashed lines),~0.2 (red solid lines) and 0.4 (green wide dashed lines) quark matter  in quark-meson model.  The values of $\sigma_{x}$ and $\sigma_{y}$ in the vacuum approximately are 92.4 MeV and 94.5~MeV, respectively.}
\end{figure}

In the   2+1 flavor quark-meson model,  the chiral condensates of both  light quarks and strange quarks  can be regarded as  the order parameters to analyze the  feature  of the chiral  phase transition.
The anisotropy parameters we work here are artifically taken as $\xi=-0.4,~0,~0.2,~0.4$, although the value of  $\xi$ in the realistic  HICs always remains positive in sign. 
 In Fig.~\ref{Fig:Condensate}, the temperature  $T$ dependences of non-strange  chiral condensate $\sigma_{x}$  and strange chiral condensate $\sigma_{y}$ for  both  isotropic and anisotropic  quark matter  at vanishing quark chemical potential are plotted.
 For  $T=0$~MeV,  $\sigma_{x}^0\approx92.4~$MeV and $\sigma_{y}^0\approx94.5~$MeV.
 As can be seen,    $\sigma_{x}$ and $\sigma_{y}$ in  both  isotropic and anisotropic  quark matter  decrease continuously with increasing temperature.
  This means that  at vanishing quark chemical potential, the restoration of the chiral symmetry  for (an-)isotropic quark matter is always a crossover phase transition. 
 And the restoration of the chiral symmetry in the strange sector is always slower than that in the non-strange sector.
   As  $\xi$ increases, the values of  $\sigma_{x}$ and $\sigma_{y}$ increase and their  melting behaviors   become more  smoother.  
This shows that  an  increase of anisotropy parameter tends to delay the chiral symmetry  restoration. 

\begin{figure}
	\includegraphics[width=0.47\textwidth]{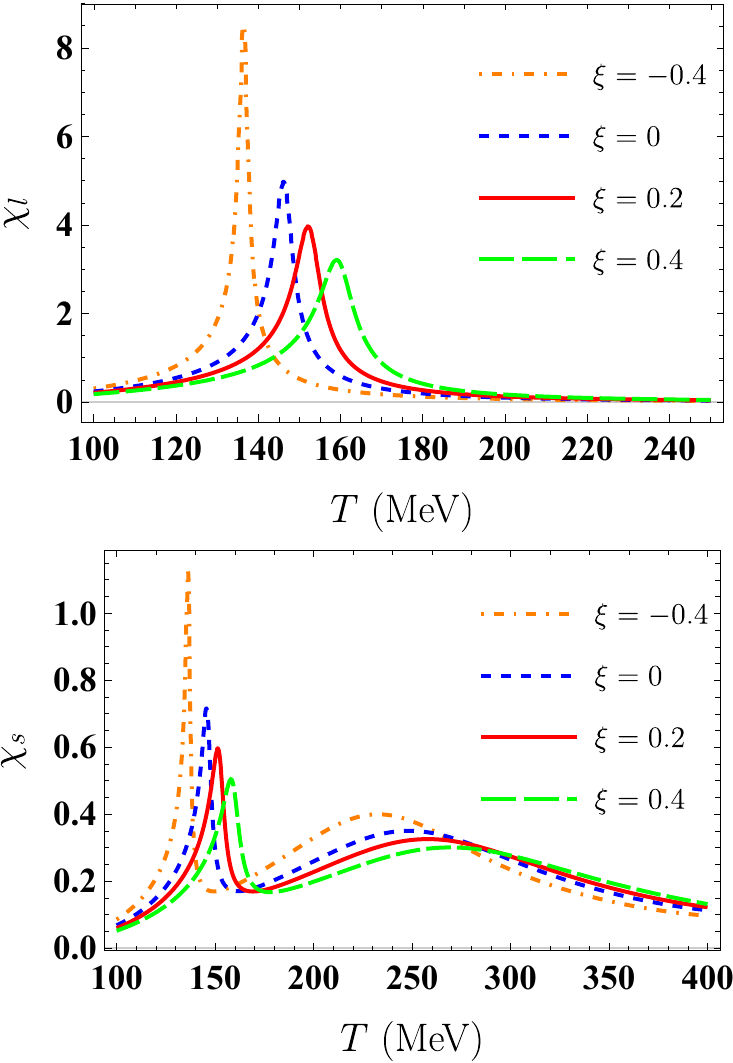}
	\caption{\label{Fig:Chi}The temperature dependences of  the susceptibilities  in non-strange sector $\chi_{l}$   (upper panel) and in strange sector $\chi_{s}$  (lower panel) at  $\mu=0$~GeV for both  isotropic ($\xi=0$ (blue dashed line))  and anisotropic (i.e., $\xi=$ $-0.4$ (orange dotted-dashed line),~0.2 (red solid line)) and 0.4 (green wide dashed line) quark matter  in the quark-meson model.}
\end{figure}

In order to  obtain the chiral critical temperature, we introduce   the susceptibilities of light quarks $\chi_{l}$ and strange quarks $\chi_{s}$, which are defined as
\begin{eqnarray}
\chi_{l}=-\frac{\partial\sigma_{x}}{\partial T}, \ \ \ \ \ \chi_{s}=-\frac{\partial\sigma_{y}}{\partial T}.
\end{eqnarray}
The thermal behaviors of both  $\chi_{l}$ and $\chi_{s}$ are presented in  Fig.~\ref{Fig:Chi}.
 We can see that  $\chi_{l}$ and $\chi_{s}$  are    peaking  up at the  particular temperatures.
 The peak position of $\chi_{l}$  determines  the   critical temperature $T^{\chi}_{c}$ for the chiral transition in  non-strange sector. Different to $\chi_{l}$,  $\chi_{s}$  have two peaks in the entire temperature domain of interest.  The temperature coordinate of  the first peak of  $\chi_{s}$  is almost  same as that  of  $\chi_{l}$,   the location of the  second broad peak of $\chi_{s}$ determines the  critical temperature  for the chiral transition of strange sector $T^{\chi}_{s}$. The chiral  critical temperature $T_{c}^{\chi}$ at vanishing quark  chemical potential is the origin of the crossover phase transition in the QCD chiral phase diagram. Furthermore, these chiral  critical temperatures are sensitive to the  variation of $\xi$. As  $\xi$ increases,    $T^{\chi}_{c,s}$    shifts towards higher temperatures as well as  the height of  ${\chi}_{l,s}$   decreases.
The exact values of both $T^{\chi}_{c}$  and    $T^{\chi}_{s}$ for different anisotropy parameters are listed in Table~\ref{tb2}.  Compared to the case of $\xi=0$, the chiral critical temperatures  $T^{\chi}_c$ and $T^{\chi}_s$   decrease by approximately $6\%$ for the case of $\xi=-0.4$.
For the cases of $\xi=0.2$ and 0.4, both  $T^{\chi}_c$ and $T^{\chi}_s$  increase by  approximately
$4\%$ and $9\%$, respectively.

\begin{figure}
\includegraphics[width=0.47\textwidth]{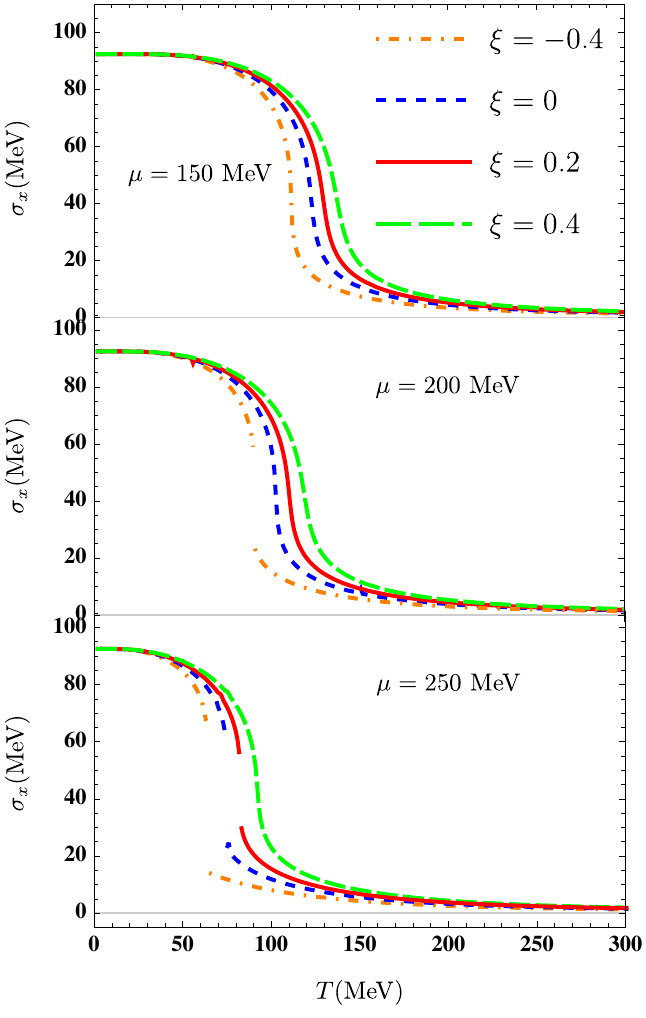}
\caption{\label{Fig:condensate-u}  The temperature dependence of the non-strange chiral condensate at $\mu=150~\mathrm{MeV}$ (upper panel), $\mu=200~\mathrm{MeV}$ (middle panel) and $\mu=250~\mathrm{MeV}$ (lower panel) in quark matter  with different anisotropy parameters, i.e., $\xi=$ $-0.4$ (orange dotted-dashed lines),~0.0 (blue dash lines),~0.2 (red solid lines) and 0.4 (green wide dashed lines)}
\end{figure}

\begin{figure}
 	\includegraphics[width=0.47\textwidth]{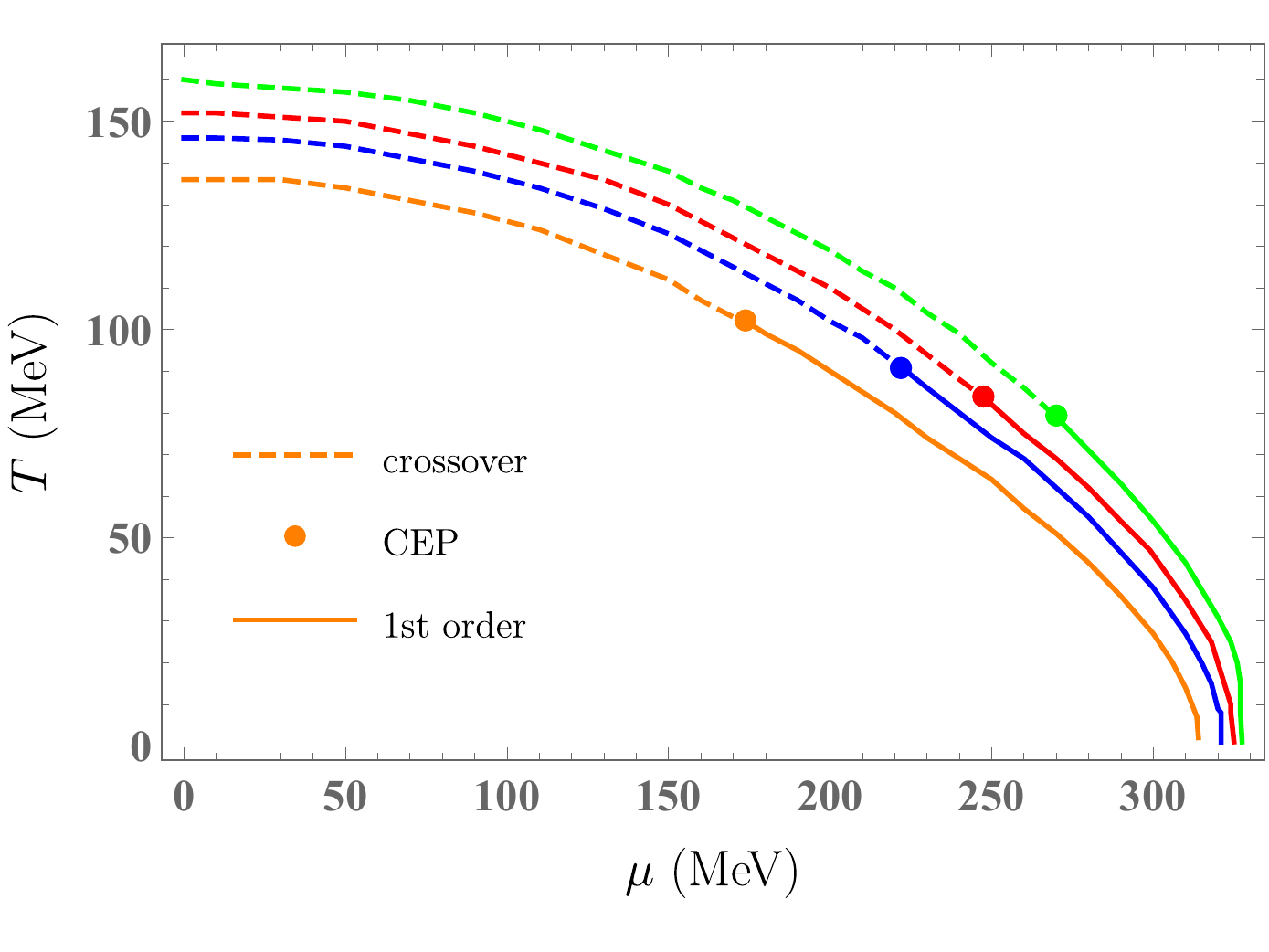}
	\caption{\label{Fig:CEP} Chiral phase diagram for different anisotropy parameters  in the quark-meson model. The solid lines represent the first-order phase transition curves, the dashed lines denote the crossover transition curves, and  the solid dots represent the positions of the CEP ($\mu_{CEP},~T_{CEP}$).  }
\end{figure}

\begin{figure*}
	\includegraphics[width=7.in,height=2.in]{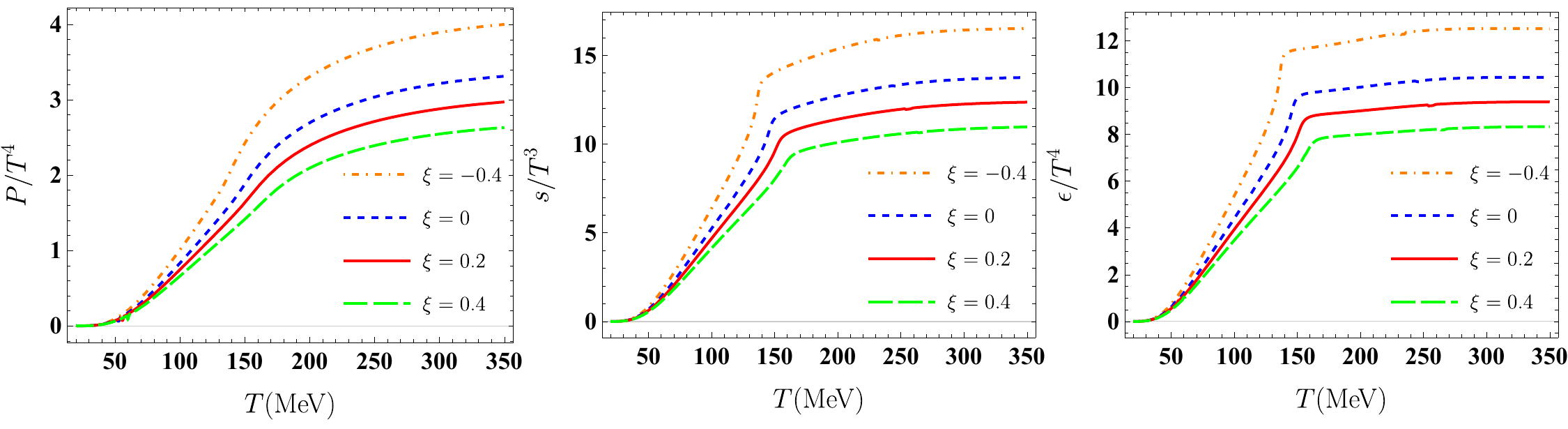}
	\caption{\label{fig:thermodynamics}  The temperature dependences of the scaled pressure $P/T^4$(left panel), scaled entropy density $s/T^3$(middle panel), and scaled energy density $\epsilon/T^4$ (right panel) for $\mu=0~\mathrm{MeV}$ in quark matter  with different anisotropy parameters, i.e., $\xi=$ $-0.4$ (orange dotted-dashed lines),~0.0 (blue dash lines),~0.2 (red solid lines) and 0.4 (green wide dashed lines).}
\end{figure*}

Next, we  extend our exploration to finite quark  chemical potential for analyzing the effect of momentum anisotropy on the structure of  QCD phase diagram.
In Fig.~\ref{Fig:condensate-u}, the temperature dependence of non-strange chiral condensate $\sigma_{x}$ for both isotropic and anisotropic quark matter  at different quark chemical potentials ($viz$., $\mu=$150~MeV,~200~MeV,~250~MeV) is plotted. At $\mu=150~$MeV, the chiral symmetry restoration with different $\xi$ still  takes place  as the crossover phase transition. 
For $\mu=200~$MeV,  the value of  $\sigma_{x}$ in the anisotropic quark matter  with $\xi=-0.4$ drops  from 60~MeV to 23~MeV, and associated susceptibility  presents a divergent behavior  at $T=90~$MeV, which signals the appearance  of a  first-order phase transition. For  $\mu=250~$MeV, the discontinuity of $\sigma_{x}$ (i.e., the first-order phase transition) also occurs at  $\xi=-0.4,~0$ and 0.2, whereas, at  $\xi=0.4$  the phase transition is still a smooth  crossover.
Thus,   for the anisotropic  matter with $\xi=0.4$, a first-order phase transition happens at  higher quark chemical potential. 
Accordingly, the  chiral phase transition diagram can be studied by outlining the location of $T^{\chi}_{c}$ for a wide range of quark chemical potential. 
And  the first-order   phase transition has to end and then changes into a crossover  is the QCD critical endpoint (CEP), at which the phase transition is of second order.
In Fig.~\ref{Fig:CEP}, the 2+1 flavor chiral phase diagram in the ($\mu$, $T$)-plane  for the quark-meson model within the effect of momentum-space anisotropy is presented. 
Along the first-order phase transition line (crossover phase transition  line), the chiral critical temperature rises from zero up to the CEP temperature  (from the $T_{CEP}$  up to the $T_{c}^{\chi}(\mu=0)$), whereas the critical quark chemical potential decreases from  the  $\mu_{c}(T=0)$ to the  $\mu_{CEP}$ (from the  $\mu_{CEP}$ to zero).
 We observe that   the phase boundary in the ($\mu$, $T$)-plane   of the quark-meson model phase diagram is shifted to higher values of $\mu$ and $T$, with increasing anisotropy parameter.
 We also  can clearly see that the position of the CEP  significantly depends on   the  variation  of  momentum  anisotropy parameter. As $\xi$ increases, the location of  the CEP shifts to higher $\mu$ and smaller $T$ domain, which is similar to the study of non-extensive effect in linear sigma model~\cite{Shen:2017etj}.
 Similar phenomenon is also observed in the literature for analyzing   the  finite size   effects  on chiral phase transition~\cite{Tripolt:2013zfa,Li:2017zny,Palhares:2009tf,Magdy:2015eda,Zhao:2019ruc}. In  Ref.~\cite{Tripolt:2013zfa}, when the system size is reduced to 4 fm, the CEP  in the quark-meson model vanishes and the whole  chiral phase boundary becomes a crossover curve.  Based on this result, we deduce that as $\xi$ increases further, the CEP may disappear.
  In this work,  for $\xi=-0.4,~0,~0.2,~0.4$, the location of the CEP is at $(T_{CEP},~\mu_{CEP})=(100,~174)$~MeV, (91,~222)~MeV,~ (84,~247)~MeV and  (79,~270)~MeV, respectively.  The value of $\mu_{CEP}$ from $\xi=-0.4$ to $\xi=0.4$ increases by about 50\%, whereas the value of $T_{CEP}$   increases by about 20\%. This means  that the influence  of momentum-space anisotropy on  the  quark chemical potential coordinate of the CEP is  more prominent compared to that on the temperature  of the CEP.   An opposite trend can be found in the study of finite volume effect~\cite{Tripolt:2013zfa}, where  the temperature coordinate of the CEP in the quark-meson model  appears to be affected  more strongly by the finite volume than CEP's quark chemical potential coordinate. 
\subsection{QCD thermodynamic quantities}
Let us now  study the influence of anisotropy parameter $\xi$ on the thermodynamics at vanishing quark chemical potential.  The $T$- and $\xi$-dependent  pressure $P(T,\xi)$, which is  derived from the thermodynamic potential, can  be given as 
\begin{eqnarray}
P(T,\xi)=-\Omega(T,\xi),
\end{eqnarray}
with the vacuum normalization $P(0,\xi)=0$.
The entropy density $s$ and energy density $\epsilon$ are defined as
\begin{eqnarray}\label{sq}
s(T,\xi)=-\frac{\partial\Omega(T,\xi)}{\partial T}
\end{eqnarray}
and
\begin{eqnarray}
 \epsilon(T,\xi)=-P(T,\xi)+Ts(T,\xi),
\end{eqnarray}
respectively.

 In Fig.~\ref{fig:thermodynamics}, the variations of the scaled pressure $P/T^{4}$, scaled entropy density  $s/T^{3}$, and  scaled energy density $\epsilon/T^{4}$     with respect to temperature  in  the quark-meson model for both  isotropic and anisotropic quark matter  are presented. As can be seen  that  the thermal behaviors of   $P/T^{4}$, $s/T^{3}$, and  $\epsilon/T^{4}$    for the anisotropic quark matter    is in agreement with  those  for the  isotropic system.
 To be specific, with increasing temperature, $P/T^{4}$,  $s/T^{3}$, and  $\epsilon/T^{4}$   first rise rapidly then tend towards a saturation value.
 At high enough temperature, the  limit values of  $P/T^4$,  $s/T^3$, and $\epsilon/T^{4}$  in the case of $\xi=-0.4$ stabilize approximately at  4.0, 16.5,~12.5, respectively, although  all these values are lower than their respective QCD Stefan-Boltzmann (SB)  limit values: $\frac{P_{SB}}{T^4}=(N_{c}^2-1)\frac{\pi^2}{45}+N_{c}N_{f}\frac{2\pi^2}{180}\simeq5.2$, $\frac{s_{SB}}{T^3}=\frac{4P_{SB}}{T^4}\simeq20.8,~\frac{\epsilon_{SB}}{T^4}=\frac{3P_{SB}}{T^4}\simeq15.6$. 
From Fig.~\ref{fig:thermodynamics}  we also can see that  the limit values of these thermodynamics at high enough temperature  still are decreasing functions of $\xi$, which  is opposite to their qualitative behaviors with  the   non-extensive parameter $q$. In Ref.~\cite{Zhao:2020xob}, at  high temperature, the limit values of  these scaled thermodynamics increase as  $q$ increases. 
 	 Moreover, their features  with $\xi$ are significantly different to those with finite volume effect. For example, Refs.~\cite{Magdy:2015eda,Bhattacharyya:2012rp} have indicated that  with increasing temperature,   $P/T^4$ first   decreases with increasing volume and then quickly saturates to the infinite volume value, in other word, these thermodynamics are insensitive to volume changes in high temperature domain.

\begin{figure}
	\includegraphics[width=0.47\textwidth]{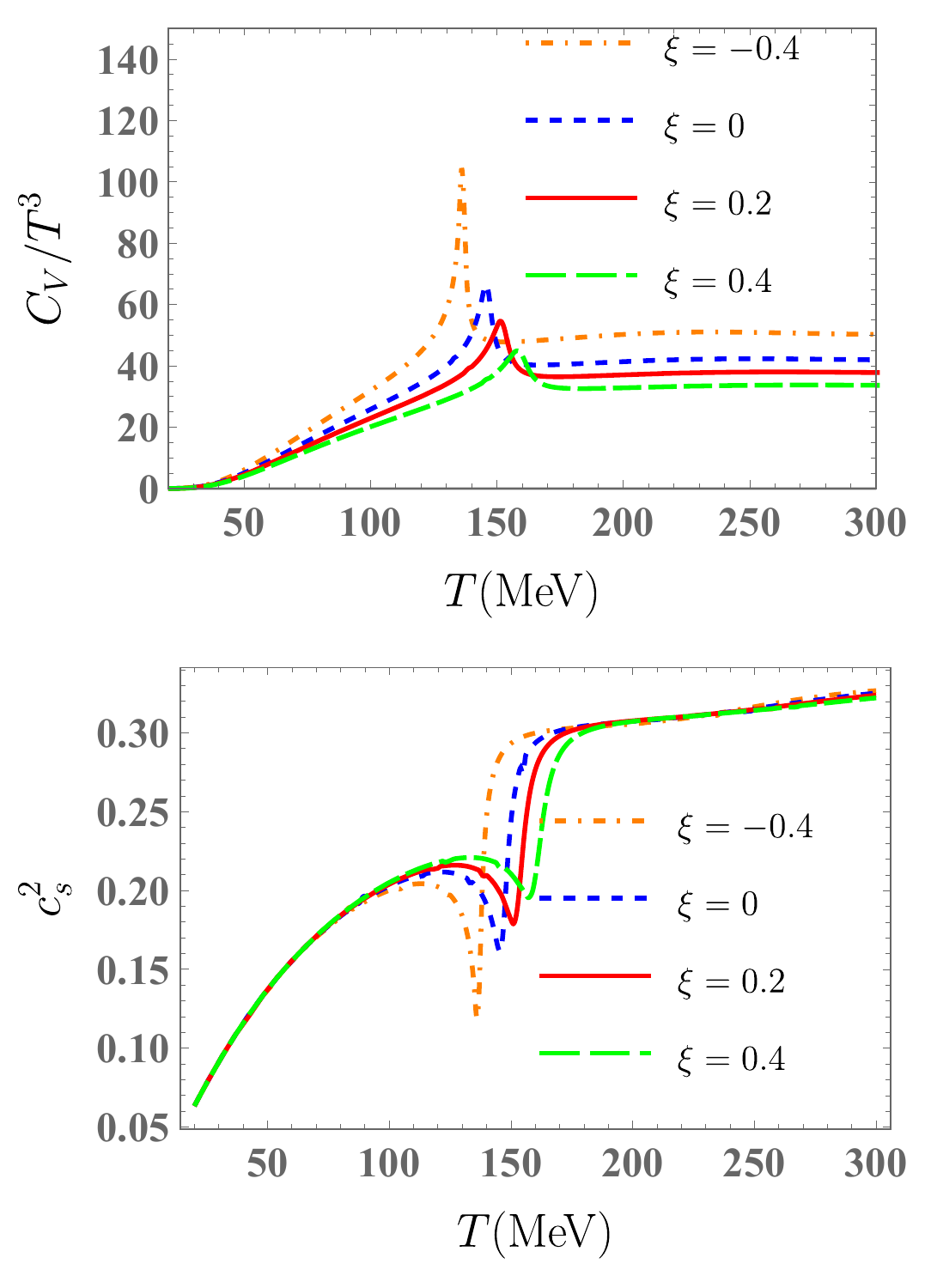}
	\caption{\label{Fig:CV-Cs}  The temperature  dependences of the scaled specific heat $C_{V}/T^3$ (upper panel)  and  the squared speed of sound $c_{s}^2$ (lower panel) at $\mu=0$~MeV for   isotropic ($\xi=0$ (blue dashed lines))  and anisotropic (i.e., $\xi=$ $-0.4$ (orange dotted-dashed lines),~0.2 (red solid lines) and 0.4 (green wide dashed lines) quark matter  in the quark-meson model.}
\end{figure}

The speed of sound squared  $c_{s}^2$  as an important quantity  in the HICs is also  studied in present work. It is defined by
\begin{eqnarray}
c_{s}^2(T,\xi)=\frac{\partial P}{\partial \epsilon}\bigg|_{V}=\frac{\partial P}{\partial T}\bigg|_{V}\bigg/\frac{\partial\epsilon}{\partial T}\bigg|_{V}=\frac{s}{C_{V}},
\end{eqnarray}
with the specific heat at constant volume $V$
\begin{eqnarray}
C_{V}(T,\xi)=\frac{\partial \epsilon}{\partial T}\bigg|_{V}=-T\frac{\partial^2\Omega}{\partial T^2}\bigg|_{V}.
\end{eqnarray}
As shown in  the   upper panel of  Fig.~\ref{Fig:CV-Cs}, the scaled specific heat  $C_{V}/T^{3}$ first rises rapidly  with increasing temperature,  reaches the maximum near the chiral critical temperature $T^{\chi}_c$, then decreases and  eventually remains constant.  Similar to  $P/T^{4}$, $s/T^{3}$, and $\epsilon/T^{4}$,  the  limit value of $C_{V}/T^{3}$  at high temperature also decreases as $\xi$ increases.
The peak of $C_{V}/T^{3}$  decreases as $\xi$ increases,  in other word, as $\xi$ increases, the critical behavior of $C_{V}/T^{3}$ is smoothed out. 
 From the lower panel of  Fig.~\ref{Fig:CV-Cs} we see that  the thermal  behavior of the  speed of sound squared   $c_{s}^{2}$ for $\xi=-0.4$ exhibits a  sharp drop near the corresponding chiral  critical temperature $T_{c}^{\chi}$,  then increases rapidly up to   the ideal gas value of $1/3$.
  Moreover, as $\xi$ increases, the dip structure of  $c_{s}^{2}$  is gradually weakened and the location of its minimum shifts to higher temperatures, which is qualitatively similar to $C_V/T^3$. At high temperature, we can see that $c_{s}^2$ is nearly unaffected by $\xi$, which is  due to that the reduction in entropy density and the increment in inverse specific heat  almost cancel each other out.
   The literature for the studies of  finite-size effect~\cite{Bhattacharyya:2012rp,Saha:2017xjq} and non-extensive effect~\cite{Zhao:2020xob} in PNJL model also has indicated that  as the system size $L$ (non-extensive parameter $q$) decreases (increases), the critical behavior of $c_{s}^2$ gradually dilutes  and    even  vanishes.  Therefore, these results of thermodynamics again emphasize that the increase of $\xi$ can hinder the restoration of  chiral symmetry.

\subsection{  Meson mass }
In this part, we study the chiral structures of scalar ($J^P=0^+$)  and pseudoscalar  ($J^P=0^-$) meson masses  at vanishing  quark chemical potential.
A detailed procedure for  calculating meson masses at finite temperature and quark chemical potential in the  quark-meson model can be found in Ref.~\cite{Schaefer:2008hk}. Here, we just sketch the outline of the related  computation.
 In quantum field theory, the scalar and pseudoscalar  meson masses generally  can be obtained from   the second derivative of the temperature- and quark chemical potential-dependent thermodynamic   potential  density $\Omega(T,\mu_{f})$ with respect to corresponding the scalar fields  
$\alpha_{S,a}=\sigma_{a}$ and  the pseudoscalar fields $\alpha_{P,a}=\pi_{a} 
(a=0,...,8)$, which can be expressed as~\cite{Schaefer:2008hk}
\begin{equation}\label{eq:m^2_{i,ab}}
	m_{i,ab}^2=\frac{\partial^2\Omega(T,\mu_{f})}{\partial\alpha_{i,a}\partial\alpha_{i,b}}\bigg|_{\mathrm{min}}=(m_{i,ab}^{\mathrm{M}})^2+(m_{i,ab}^{\mathrm{T}})^2
\end{equation}
where  the subscript $i=S(P)$ denotes the scalar (pseudoscalar) mesons.
The first term on the right-hand side of Eq.~(\ref{eq:m^2_{i,ab}}) is  vacuum  mass squared matrices calculated from the second derivative of purely mesonic potential. The second term represents  the modification  of   mass squared matrices due to fermionic thermal correction at finite temperature and quark chemical potential, which in an anisotropic system can be written  as 
\begin{widetext}
	\begin{equation}\label{eq:mass}
	\begin{aligned}
(\delta m_{i,ab}^{T})^2&=	\frac{\partial\Omega_{q\bar{q}}(T,\mu_{f},\xi)}{\partial\alpha_{i,a}\partial\alpha_{i,b}}\\
	&=2N_{c}\sum_{f=l,s}\int\frac{dp}{4\pi^2}\frac{p^2}{E_{f}}\left\{\left[f_{q,f}^0\left(m^2_{f,ab}-\frac{m^2_{f,a}m_{f,b}^2}{2E_{f}^2}\right)-\frac{f_{q,f}^0(1-f_{q,f}^0)}{2E_{f}T}m_{f,a}^2m_{f,b}^2\right]\right.\\
	&\phantom{=\;\;}\left.\times\left[1-\frac{\xi p^2}{6E_{f}T}(1-f_{q,f}^0+\frac{T}{E_{f}})\right]+\frac{\xi p^2f_{q,f}^0}{12E_{f}^2T^2}m_{f,a}^2m_{f,b}^2\left[\frac{2T^2}{E_{f}^2}+\frac{T}{E_{f}}-\frac{Tf^0_{q,f}}{E_{f}}-f_{q,f}^0(1-f_{q,f}^0)\right]+q\rightarrow\bar{q}\right\} .
	\end{aligned}\\
	\end{equation}
	\end{widetext}
The squared constituent quark mass derivative with respect  to  meson field $\partial m_{f}^2/\partial\alpha_{i,a}\equiv m_{f,a}^2$, and that with respect to meson fields $\partial^2 m_{f}^2/(\partial\alpha_{i,a}\partial\alpha_{i,b})\equiv m_{f,ab}^2$  for different flavors are listed in  the Table III of  Ref.~\cite{Schaefer:2008hk}.
When $\xi=0$, Eq.~(\ref{eq:mass})  can reduce to the result for an isotropic system.
 Thereafter, the  squared masses of four scalar meson states are given as~\cite{Tawfik:2014gga,Schaefer:2008hk,Lenaghan:2000ey} 
\begin{eqnarray}
m_{a_{0}}^2&=&(m_{a_{0}}^\mathrm{M})^2+(\delta m_{S,11}^T)^2,\label{eq:mao}
\\
m_{\kappa}^2&=&(m_{\kappa}^\mathrm{M})^2+(\delta m_{S,44}^T)^2,
\\
m_{\sigma}^2&=&m_{S,00}^2\cos^2\theta_{S}+m_{S,88}^2\sin^2\theta_{S}\nonumber\\
&&+2m_{S,08}^2\sin\theta_{S}\cos\theta_{S},\\
m_{f_{0}}^2&=&m_{S,00}^2\sin^2\theta_{S}+m_{S,88}^2\cos^2\theta_{S}\nonumber\\
&&-2m_{S,08}^2\sin\theta_{S}\cos\theta_{S}.
\end{eqnarray}
And the four pseudoscalar meson masses are
\begin{eqnarray}
m_{\pi}^2 &=& (m_{\pi}^{\mathrm{M}})^2+(\delta m_{P,11}^T)^2,\\
m_{K}^2&=&(m_{K}^{\mathrm{M}})^2+(\delta m_{P,44}^T)^2,\\
m_{\eta^{'}}^2&=& m_{P,00}^2\cos^2\theta_{P}+m_{P,88}^2\sin^2\theta_{P}\nonumber\\&&+2
m_{P,08}^2\sin\theta_{P}\cos\theta_{P},\\
m_{\eta}^2&=& m_{P,00}^2\sin^2\theta_{P}+m_{P,88}^2\cos^2\theta_{P}\nonumber\\&&-2
m_{P,08}^2\sin\theta_{P}\cos\theta_{P},\label{eq:meta}
\end{eqnarray}
where the mixing angles $\theta_{S(P)}$ read as 
\begin{eqnarray}
\tan2\theta_{i}=(\frac{2m_{i,08}^2}{m_{i,00}^2-m_{i,88}^2}),\quad i=S,P.
\end{eqnarray}
and $m_{i,00/88/08}^2=(m_{i,00/88/08}^\mathrm{M})^2+\delta(m_{i,00/88/08}^{T})^2$.
The detailed descriptions  of the vacuum contributions   [$(m_{a_{0}}^\mathrm{M})^2$, $(m_{\kappa}^\mathrm{M})^2$, $(m_{\pi}^{\mathrm{M}})^2$, $(m_{K}^{\mathrm{M}})^2$ and $(m_{i,00/88/08}^\mathrm{M})^2$] from purely mesonic potential  in Eqs.~(\ref{eq:mao})-(\ref{eq:meta}) can be found  from Refs.~\cite{Tawfik:2014gga,Schaefer:2008hk}. 
\begin{figure*}
	\includegraphics[width=6.5in,height=7.in]{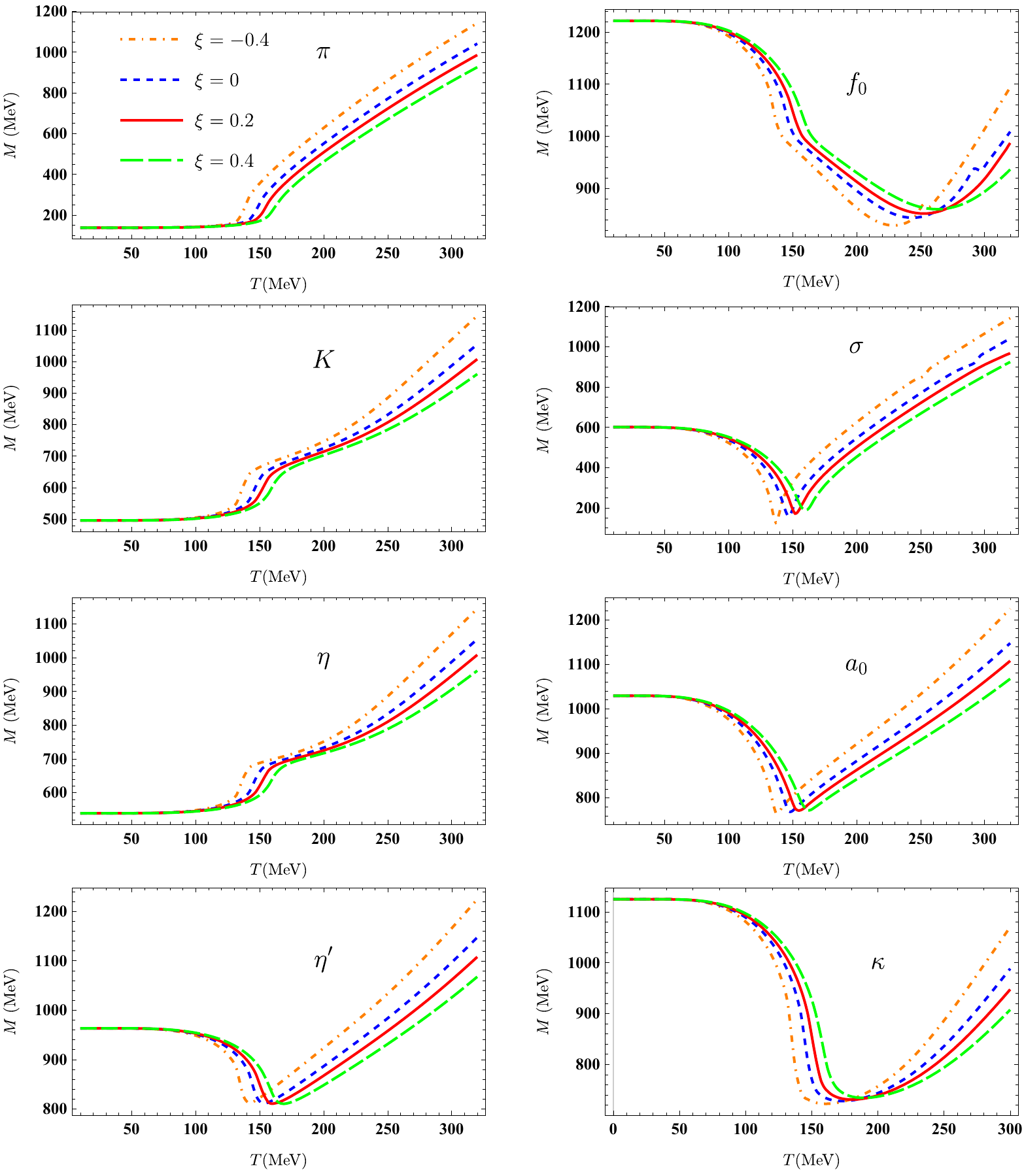}
	\caption{\label{fig:mass} The temperature dependences of the  pseudoscalar 
		mesons $\pi$,~$K$,~$\eta'$,~$\eta$ (left panels)  and scalar mesons $f_{0}$,~$\sigma$,~$a_{0}$,~$\kappa$ (right  panels) at $\mu=0~\mathrm{MeV}$   for  both  isotropic ($\xi=0$ (blue dashed lines))  and anisotropic (i.e., $\xi=$ $-0.4$ (orange dotted-dashed lines),~0.2 (red solid lines) and 0.4 (green wide dashed lines) quark matter  in the quark-meson model.}
\end{figure*}

The left panels and right panels of Fig.~\ref{fig:mass}  display  the $T$-dependent masses of the pseudoscalar ($\pi$,~$K$,~$\eta'$,~$\eta$) and scalar ($f_{0}$,~$\sigma$,~$a_{0}$,~$\kappa$) mesons for both isotropic and anisotropic quark matter  in  the  quark-meson model, respectively.  
We can see that for a fixed anisotropy parameter, the masses of the pseudoscalar meson sectors  $\pi$, $K$, and $\eta$  remain constant up  to near the chiral critical temperature of non-strange condensate $T_{c}^{\chi}$, whereas  the masses of  $\eta'$ and scalar meson sectors $\sigma$, $a_{0}$, $\kappa$ remain constant at low temperature and then decreases  before reaching $T_{c}^{\chi}$.   For the pesudoscalar meson sector $f_{0}$, its mass  also remains constant at low temperautre but  decreases before reaching the chiral critical temperature of   strange condensate $T_{s}^{\chi}$.
For  pseudoscalar meson sectors $\pi$, $K$ and $\eta$, their masses always decrease with increasing $\xi$ at $T>140$~MeV.
 However, for $\eta'$ and pseudoscalar meson sectors ($\pi$,~$K$,~$\eta'$,~$\eta$), the dependence of their  masses on anisotropy parameter $\xi$ is nonmonotonic in the entire temperature domain of interest.  More exact, with the increase of $\xi$,  the masses of $\eta'$ ~$\sigma$,~$a_{0}$,~$\kappa$  first increase in  low temperature domain ($100~\mathrm{MeV}<T<160$~MeV),  then  decrease  in higher temperature domain  ($T>160$~MeV).  For $f_{0}$,   its mass increases  with increasing $\xi$ at $T<270$~MeV ($viz$., $T_{s}^{\chi}(\xi=0.4)$) and  decreases afterward.
 As a whole, near above  $T_{c}^{\chi}$ or  $T_{s}^{\chi}$, all  mesons  become unphysical degrees of freedom and their masses become degenerate,   which signals the restoration of   chiral symmetry. 
 In Fig.~\ref{fig:mass} we  can also see  that  with the increase of $\xi$, the temperature coordinate at which meson masses begin to degenerate can be shifted to higher temperatures.  This again shows that an increase of momentum-space anisotropy parameter can  hinder the restoration of  chiral symmetry.
The qualitative behaviors of these meson masses with $\xi$ are different to the results for  analyzing the finite size dependence of meson masses within PNJL model~\cite{Ya-Peng:2018gkz,Bhattacharyya:2012rp}, where $K$, $\eta$, and $\eta'$  have a significant volume dependence in lower temperature domain ($T<100~$MeV).
\subsection{Transport coefficient}

Studying  transport properties  is essential to  deeply understand the dynamical evolution of the strongly interacting matter. In this part,  we discuss the influence of momentum-space anisotropy on  transport coefficients, such as shear viscosity $\eta$, electrical conductivity $\sigma_{el}$, and  bulk viscosity   $\zeta$ in  quark matter. Due to the effect of  momentum-space anisotropy is encoded in  the parton distribution functions, the general expressions of these transport coefficients, which are obtained by solving the relativistic Boltzmann equation  in  relaxation time approximation,  need  to do some modifications~\cite{Rath:2019vvi,Zhang:2020efz,Thakur:2017hfc,Srivastava:2015via}. Therefore, using the results in Refs.~\cite{Rath:2019vvi,Zhang:2020efz},  the formulas of $\xi$-dependent transport coefficients at zero quark chemical potential are given as
\begin{widetext}
	\begin{eqnarray}\label{eq:shear}
	\eta_{}=\sum_{f}\frac{d_f}{15T}\int \frac{dp}{\pi^2}\frac{p^6}{E_{f}^2}\left[\tau_{q,f}f_{q,f}^0(1-f_{q,f}^0)\right]-\sum_{f}\frac{\xi d_{f}}{90T^2}\int \frac{dp}{\pi^2}\frac{p^8}{E_{f}^3}[\tau_{q,f}f_{q,f}^0(1-f_{q,f}^0)(1-2f_{q,f}^0+\frac{T}{E_{f}})],
	\end{eqnarray}
	\begin{eqnarray}\label{eq:electrical}
	\sigma_{el}=\sum_{f}\frac{d_{f}q_{f}^2}{3T}\int \frac{dp}{\pi^2}\frac{p^4}{E_{f}^2}[\tau_{q,f}f_{q,f}^0(1-f_{q,f}^0)](1+\frac{\xi}{3})-\sum_{f}\frac{q_{f}^2\xi d_f}{18T^2}\int\frac{dp}{(2\pi)^3}\frac{p^6}{E_{f}^3}[f_{q,f}^0(1+f_{q,f}^0)(1-2f_{q,f}^0+\frac{T}{E_{f}})],
	\end{eqnarray}
	\begin{eqnarray}\label{eq:bulk}
	\zeta_{}&=&\sum_{f}\frac{d_{f}}{T}\int \frac{dp}{\pi^2}\frac{p^2}{E_{f}^2}\left[(\frac{1}{3}-c_{s}^2)p^2-c_{s}^2m_{f}^2+c_{s}^2m_{f}T\frac{dm_{f}}{dT}\right]^2[\tau_{q,f}f_{q,f}^0(1-f_{q,f}^0)]
	\nonumber\\
	&&-\sum_{f}\frac{\xi d_{f}}{6T^2}\int \frac{dp}{\pi^2} \frac{p^4}{E_{f}^3}\left[(\frac{1}{3}-c_{s}^2)p^2-c_{s}^2m_{f}^2+c_{s}^2m_{f}T\frac{dm_{f}}{dT}\right]^2[\tau_{q,f}f_{q,f}^0(1-f_{q,f}^0)(1-2f_{q,f}^0)]\nonumber\\
	&&-\sum_{f}\frac{\xi d_{f}}{6T}\int \frac{dp}{\pi^2} \frac{p^4}{E_{f}^4}\left[\frac{1}{9}p^4   -\bigg(c_{s}^2(m_{f}^2+p^2)-c_{s}^2m_{f}T\frac{dm_{f}}{dT}\bigg)^2\right]\tau_{q,f}f_{q,f}^0(1-f_{q,f}^0).
	\end{eqnarray}
\end{widetext}
Here, $d_{f}$ is the degeneracy factor  for $f$-flavor quark.
The quark electric charge $q_f$ is given explicitly by $q_u=-q_{\bar{u}}=2e/3$ and $q_{d,s}=-q_{\bar{d},\bar{s}}=-e/3$. The electron charge reads $e=(4\pi\alpha_s)^{1/2 }$ with the fine structure constant $\alpha_s\simeq1/137$.
Different to the formula of bulk viscosity in Ref.~\cite{Rath:2019vvi}, we replace the original term $[(\frac{1}{3}-c_{s}^2)p^2]^2$ in the integrand  with $\left[(\frac{1}{3}-c_{s}^2)p^2-c_{s}^2m_{f}^2+c_{s}^2m_{f}T\frac{dm_{f}}{dT}\right]^2$ to incorporate the in-medium effect. In the treatment of  the relaxation time $\tau_{q,f}$, we rougly  take a constant value  $\tau_{q,f}=1\ \mathrm{fm}$ for the computation.
In the weakly anisotropic system, the former terms in Eqs.~(\ref{eq:shear})-(\ref{eq:bulk}) are  significantly larger than the latter terms in magnitude due to the difference in momentum power of respective integrand.  Therefore,   transport coefficients are still mainly  dominated by the first term of related expressions on the quantitative.
 \begin{figure}
	\includegraphics[width=0.5\textwidth]{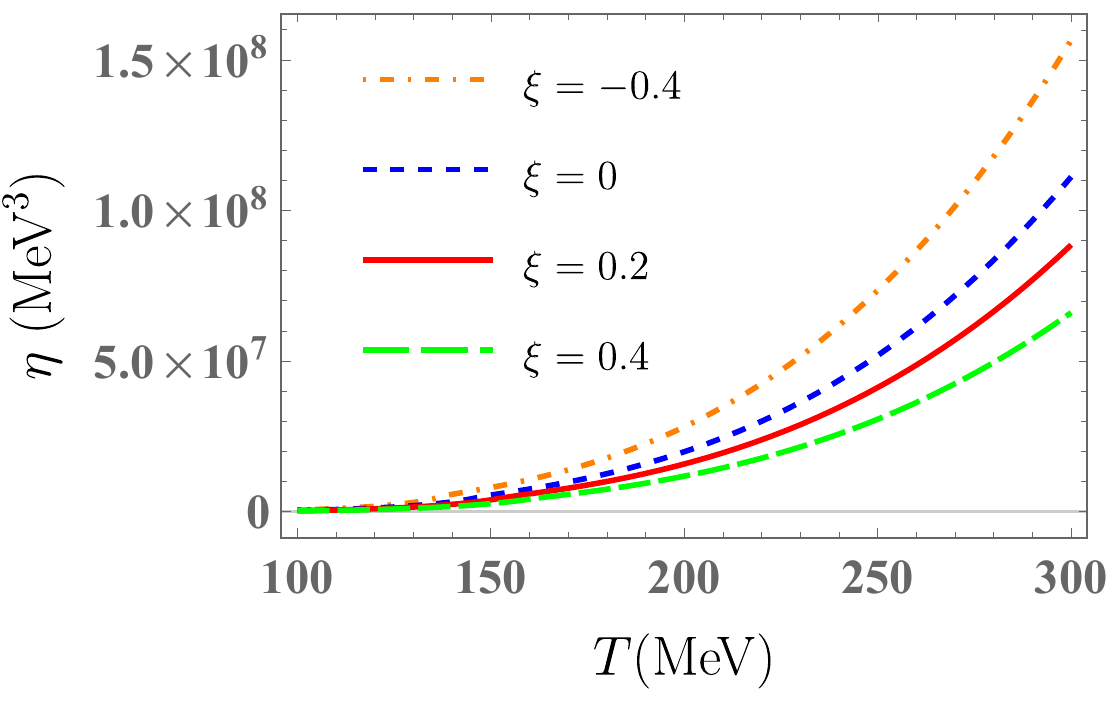}
	\caption{\label{Fig:shear}The temperature dependence of   shear viscosity $\eta$   at $\mu=0$~MeV in  quark matter with  $\xi= -0.4$ (orange dotted-dashed line), 0.0 (blue dashed line),~0.2 (red solid line), 0.4 (green wide dashed line).}
\end{figure}
\begin{figure}
	\includegraphics[width=0.45\textwidth]{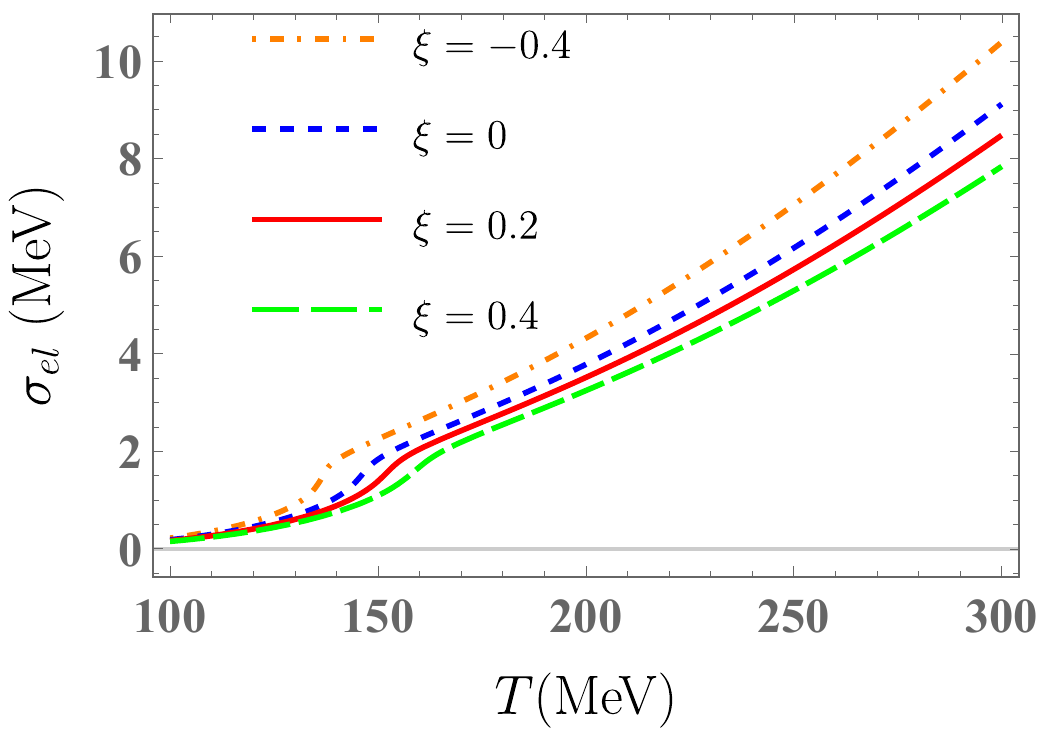}
	\caption{\label{Fig:electrical} The temperature dependence of  electrical conductivity $\sigma_{el}$  at $\mu=0$~MeV in  quark matter with  $\xi= -0.4$ (orange dotted-dashed line), 0.0 (blue dashed line),~0.2 (red solid line), 0.4 (green wide dashed line). }
\end{figure}
\begin{figure}
	\includegraphics[width=0.48\textwidth]{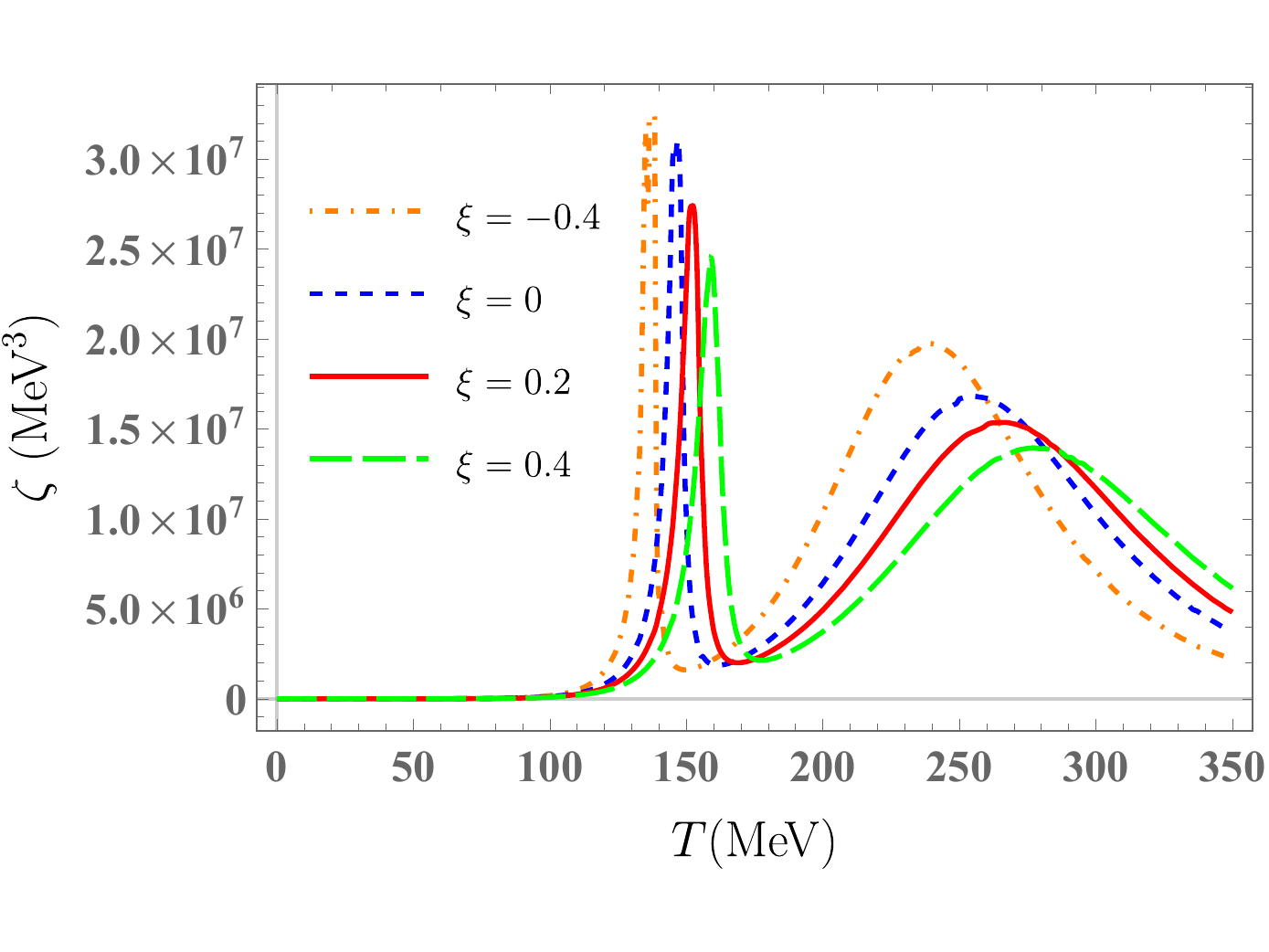}
	\caption{\label{Fig:bulk} The temperature dependence of   bulk viscosity  $\zeta$   at $\mu=0$~MeV in  quark matter with  $\xi= -0.4$ (orange dotted-dashed line), 0.0 (blue dashed line),~0.2 (red solid line), 0.4 (green wide dashed line).}
\end{figure}

The variation of shear viscosity $\eta$ with temperature    at  vanishing quark chemical potential for  both isotropic and  anisotropic quark matter  is  shown in Fig.~\ref{Fig:shear}.   We see that   $\eta$ in the (an-)isotropic quark matter  rises monotonically with increasing temperature because the $T$ dependence of $\eta$ is mainly  coming from the  quark distribution function $f_{q,f}^0$ in the associated integrand.  
For the qualitative behavior of $\eta $ with $\xi$, we also can well understand from the associated expression. 
 In the vicinity of the chiral critical temperature $T_{c}^{\chi}$, $\eta$ slightly decreases as $\xi$ grows due to decreasing behavior of the Boltzmann factor $e^{-m_{f}(T)/T}$ with $\xi$. In  higher temperature  domain ($T>{\tiny }160$~MeV), the decreasing feature of $\eta$ is negligible  due to the unsensitivity of the constituent quark masses to $\xi$. 
However, the absolute value of the second term in Eq.~(\ref{eq:shear}) significantly increases with an increase of  $\xi$. As a result, $\eta$ decreases as $\xi$ grows. This  is similar to the  result in   Ref.~\cite{Rath:2019vvi}, where   $\eta$ for the QGP is calculated  in quasiparticle model.
 For  electrical conductivity  $\sigma_{el}$, its thermal behavior is similar to $\eta$, the quantitive difference between $\eta$ and $\sigma_{el}$ is mainly coming from the different momentum power of  respective integrand.
 Similar  to  shear viscosity, the $\xi$ dependence of $\sigma_{el}$ is also determined by the second term in the associated expression.
In Fig.~\ref{Fig:electrical},
  we observe that  $\sigma_{el}$ decreases as $\xi$ increases,
 which is also  qualitatively  consistent  with the results  of $\sigma_{el}$ for the QGP in quasiparticle model~\cite{Thakur:2017hfc,Srivastava:2015via}. 
 The dependences of $\eta$ and $\sigma_{el}$ on momentum-space anisotropy are different from those on finite system size $L$ in the framework of (P)NJL model.  In Ref.~\cite{Saha:2017xjq},
 both $\eta$ and $\sigma_{el}$ first increase as  $L$ decreases in low temperature domain, whereas,  the size effect  nearly vanishes in    high temperature domain. Furthermore, the result in  Ref.~\cite{Zhao:2020xob}  has indicated that  both  $\eta$ and $\sigma_{el}$ in PNJL model also increase obviously  as non-extensive parameter $q$ increases at $T>150$~MeV.

Next, we discuss the temperature dependence of   bulk viscosity $\zeta$ at zero quark chemical potential for both isotropic and anisotropic quark matter.
As shown in Fig.~\ref{Fig:bulk},  for a fixed anisotropy parameter, $\zeta$ is   peaking up in the vicinities of both  $T^{\chi}_{c}$ and   $T^{\chi}_{s}$, which is significantly different to the thermal behaviors of $\eta$ and $\sigma_{el}$.  
We also note that  the thermal profile of $\zeta$ is similar to  $dm_{s}/dT$ or $\chi_{s}$, which may be attributed  that   the qualitative behavior of $\zeta$ is mainly  govern by $dm_{s}/dT$ rather than the quark distribution function in associated integrand of Eq.~(\ref{eq:bulk}).
Due to the decreasing feature of the peak of  $dm_{s}/dT$ with  increasing $\xi$,   the  double-peak structure of $\zeta$ can be weakened as $\xi$ grows and the positions of peaks shift to higher temperatures,  as  shown in Fig.~\ref{Fig:bulk}. 
 The diluting  effect  of $\xi$ on the critical behavior of  $\zeta$   is similar to the studies regarding     finite volume effect and non-extensive effect. In Ref.~\cite{Saha:2017xjq},  the double-peak structure of $\zeta$  even  converts to one broadened peak structure when the system size is reduced to $2\ \mathrm{fm}$. And in Ref.~\cite{Zhao:2020xob}, as non-extensive parameter $q$ increases to 1.1, the two peaks of $\zeta$  also  begin to merge into a broad one.
 
\section{Summary and Conclusion}\label{summary}
 In this work, an anisotropy parameter $\xi$, which reflects the degree  of momentum-space anisotropy arising from different expansion rates of the fireball generated in HICs  along longitudinal and radial direction,  for the first time, is introduced in the 2+1 flavor quark-meson model by replacing the  isotropic distribution function in the thermodynamic potential of the quark-meson model with the anisotropic one. 
 The effect of $\xi$ on the  chiral  properties, thermodynamics, meson masses, and  transport properties  in quark matter are investigated.  We  find that the chiral  phase transition of quark matter with different anisotropy parameters is  always a crossover at vanishing quark chemical potential.
 At finite quark chemical potential,  the  temperature  of the CEP is affected more significantly by the anisotropy parameter than its quark chemical potential, which is opposite to the study of finite  volume effect.
We also demonstrate that  at  high temperature, the  limit values  of various scaled thermodynamics ($P/T^4$,~$s/T^3$,~$\epsilon/T^4$,~$C_{V}/T^3$) are quite sensitive to $\xi$. As $\xi$ increases, their limit values decrease, which is different to the finite size effect but rather similar to non-extensive effect.  
And the critical behavior of $C_{V}/T^3$ and $c_{s}^2$ can be  smoothed out with increasing $\xi$.
For  scalar and pseudoscalar mesons, the temperature, where  their masses begin to degenerate, is enhanced   as $\xi$ rises,  which implies that an increase of $\xi$ can hinder the restoration of chiral symmetry.
Finally,  the transport coefficients, such as shear viscosity~$\eta$, electrical conductivity~$\sigma_{el}$, and bulk viscosity~$\zeta$  for both isotropic and anisotropic quark matter,  are also calculated. Our results show that $\eta$ and $\sigma_{el}$  rise  with increasing  temperature, while the thermal behavior of  $\zeta$ exhibits a  noticeble  double-peak structure.
It is seen that  $\eta$ and $\sigma_{el}$  decrease monotonically as $\xi$ increases, whereas the qualitative behavior of $\zeta$ with $\xi$ is similar to $\chi_{s}(\xi)$.
With increasing $\xi$, the double-peak structure of $\zeta$ can be weakened, and the positions of peaks shift to higher temperatures.

In present work, we only focus on the chiral aspect of the QCD phase diagram, the exploration of the confinement phase transition in an anisotropic quark matter also can be addressed via including the Polyakov-loop potential. In the   Polyakov-loop improved quark-meson model,  the chiral phase transition and the  location of   the CEP  will be  affected further.
For  the calculation of transport coefficients in this study, the relaxation time  of quark is  assumed to be a constant. However, in the  realistic interaction scenario, the relaxation time    may also vary with the momentum anisotropy. These issues are our future research directions.
 Moreover, note that   a spheroidal  momentum-space anisotropy specified by  one
 anisotropy parameter in one preferred propagation direction   is considered  in this work, however,  the introduction of additional anisotropy parameters is necessary to provide a better characterization of the QGP properties. 
 The   chiral and  confinement  phase  transitions  in quark matter with   ellipsoidal momentum-anisotropy ~\cite{Kasmaei:2016apv,Kasmaei:2018yrr} characterized by two independent anisotropy parameters, also can be done using  PNJL or PQM model. The works of  these directions are in progress and we expect to report our results soon.
 
\acknowledgments
This work is supported by the National Natural Science Foundation of China under Grants No.11935007.

\end{document}